%% file: main.tex
\documentclass{article}

\usepackage{microtype}
\usepackage{graphicx}
\usepackage{booktabs} 
\usepackage{float}

\usepackage[accepted]{tidecache-preprint}

\input{packages}
\input{macros}

\usepackage{fancyhdr} 
\newcommand{\preprinttitle}{\system: Efficient Hybrid LLM Serving with Transparent Cache Management and Layer-Wise Streaming}
\mlsystitlerunning{\smash{\resizebox{0.92\textwidth}{!}{\preprinttitle}}}
\hypersetup{pdfsubject={Hybrid LLM serving}}

\begin{document}

\twocolumn[
\mlsystitle{\preprinttitle}

\begin{mlsysauthorlist}
\mlsysauthor{Jinbin Luo}{mit}
\mlsysauthor{Xiaozhe Yao}{eth}
\mlsysauthor{Shang Yang}{mit}
\mlsysauthor{Zhijian Liu}{ucsd}
\mlsysauthor{Song Han}{mit}
\mlsysauthor{Qinghao Hu}{mit}
\par\smallskip
{\small
\textsuperscript{1}Massachusetts Institute of Technology\quad
\textsuperscript{2}ETH Zurich\quad
\textsuperscript{3}University of California, San Diego}
\end{mlsysauthorlist}

\mlsyskeywords{LLM serving, hybrid models, KV cache, offloading}

\vskip 0.15in

\input{sections/0-abstract}
]

\input{sections/1-intro}
\input{sections/2-background}
\input{sections/3-motivation}
\input{sections/4-method}
\input{sections/5-evaluation}
\input{sections/6-related}
\input{sections/7-conclusion}


\bibliography{refs}
\bibliographystyle{tidecache-preprint}

\input{sections/8-ai-disclosure}

\appendix
\newpage


\end{document}

%% file: packages.tex
\usepackage{amsmath,amssymb,amsfonts}
\usepackage{graphicx}
\usepackage{textcomp}
\usepackage{xcolor}
\usepackage{multirow}
\usepackage{booktabs}
\usepackage{pifont}
\usepackage{array}
\usepackage{xspace}
\usepackage{soul}
\usepackage{subcaption}
\usepackage{fancyhdr}
\usepackage{colortbl}
\usepackage{makecell}
\usepackage[bookmarks=true,breaklinks=true,letterpaper=true,colorlinks,citecolor=blue,linkcolor=blue,urlcolor=blue]{hyperref}
\usepackage{enumitem}

\usepackage{algorithm}
\usepackage{amsmath}
\usepackage{amsfonts}

\usepackage{microtype}

\usepackage[utf8]{inputenc} 
\usepackage{inconsolata} 
\usepackage[most]{tcolorbox} 
\usepackage{soul} 
\usepackage{siunitx} 
\usepackage{pifont} 
\usepackage{xurl} 
\usepackage[textsize=scriptsize, linecolor=magenta, bordercolor=magenta,
  backgroundcolor=white, textwidth=45pt]{todonotes}
\usepackage{marginnote}

\usepackage[noindentafter]{titlesec}
\titlespacing\section{0pt}{2pt}{2pt} 
\titlespacing\subsection{0pt}{2pt}{2pt}
\titlespacing\subsubsection{0pt}{2pt}{2pt}
\titlespacing\paragraph{0pt}{0ex plus .2ex minus .1ex}{1em}

\usepackage[skip=3pt, font=small]{caption}

%% file: macros.tex
\newcommand{\sect}[1]{Section~\ref{#1}}

\newcommand{\eqn}[1]{Equation~\ref{#1}}

\newcommand{\fig}[1]{Figure~\ref{#1}}

\newcommand{\tab}[1]{Table~\ref{#1}}

\newcommand{\alg}[1]{Algorithm~\ref{#1}}

\makeatletter
\DeclareRobustCommand\onedot{\futurelet\@let@token\@onedot}
\def\@onedot{\ifx\@let@token.\else.\null\fi\xspace}

\makeatother

\definecolor{mydarkblue}{rgb}{0,0.08,1}
\definecolor{mydarkgreen}{rgb}{0.02,0.6,0.02}
\definecolor{mydarkred}{rgb}{0.8,0.02,0.02}
\definecolor{mydarkorange}{rgb}{0.40,0.2,0.02}
\definecolor{mydarkgold}{rgb}{0.6, 0.4, 0.05}
\definecolor{myred}{rgb}{1.0,0.0,0.0}
\definecolor{mygold}{rgb}{0.75,0.6,0.12}
\definecolor{myblue}{rgb}{0,0.2,0.8}
\definecolor{mylightblue}{rgb}{0.827,0.937,0.945}
\definecolor{mydarkgray}{rgb}{0.66,0.66,0.66}
\definecolor{mygray}{rgb}{0.85,0.85,0.85}

\newcommand{\cmark}{\raisebox{-0.1ex}{\includegraphics[height=1.65ex]{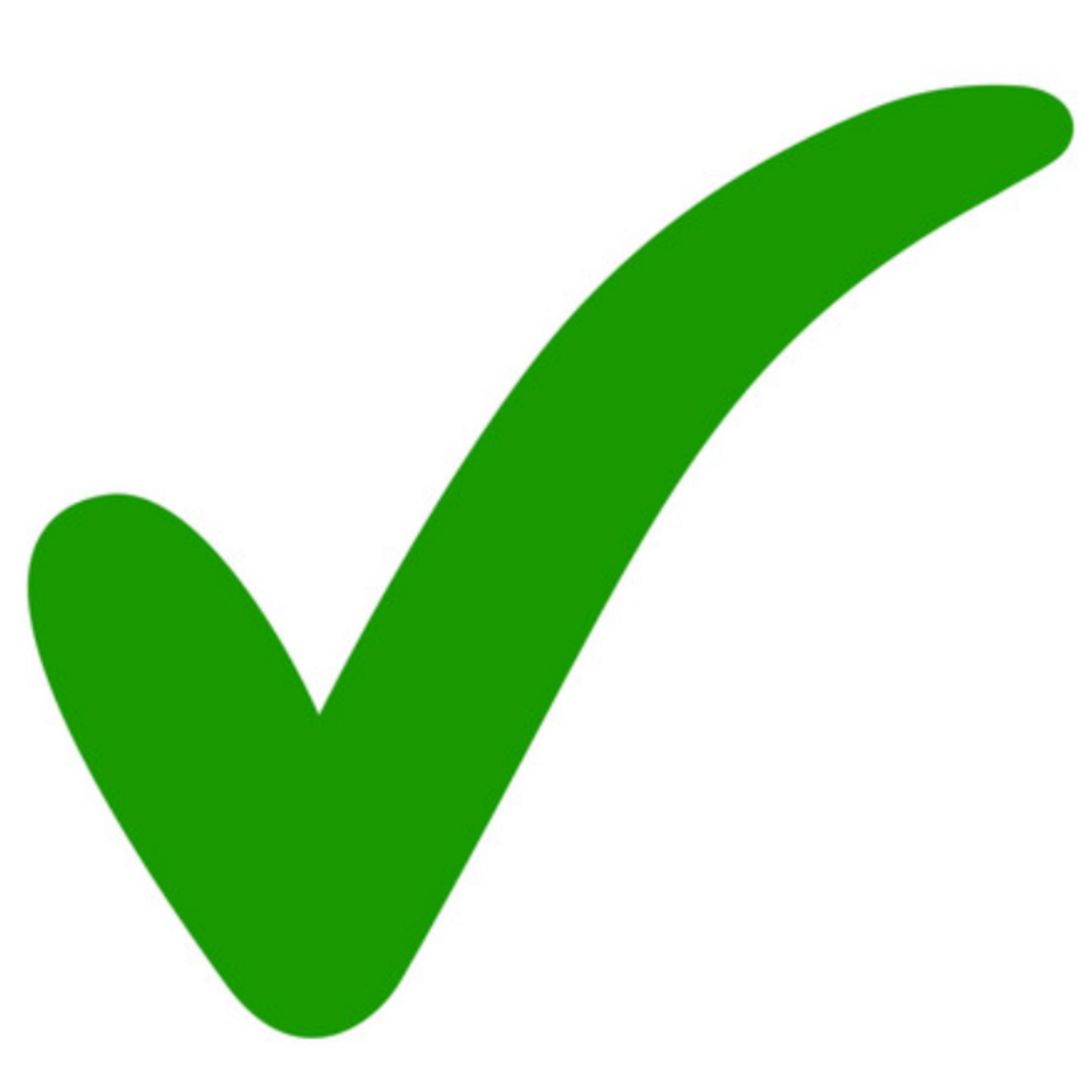}}}%
\newcommand{\xmark}{\raisebox{-0.1ex}{\includegraphics[height=1.65ex]{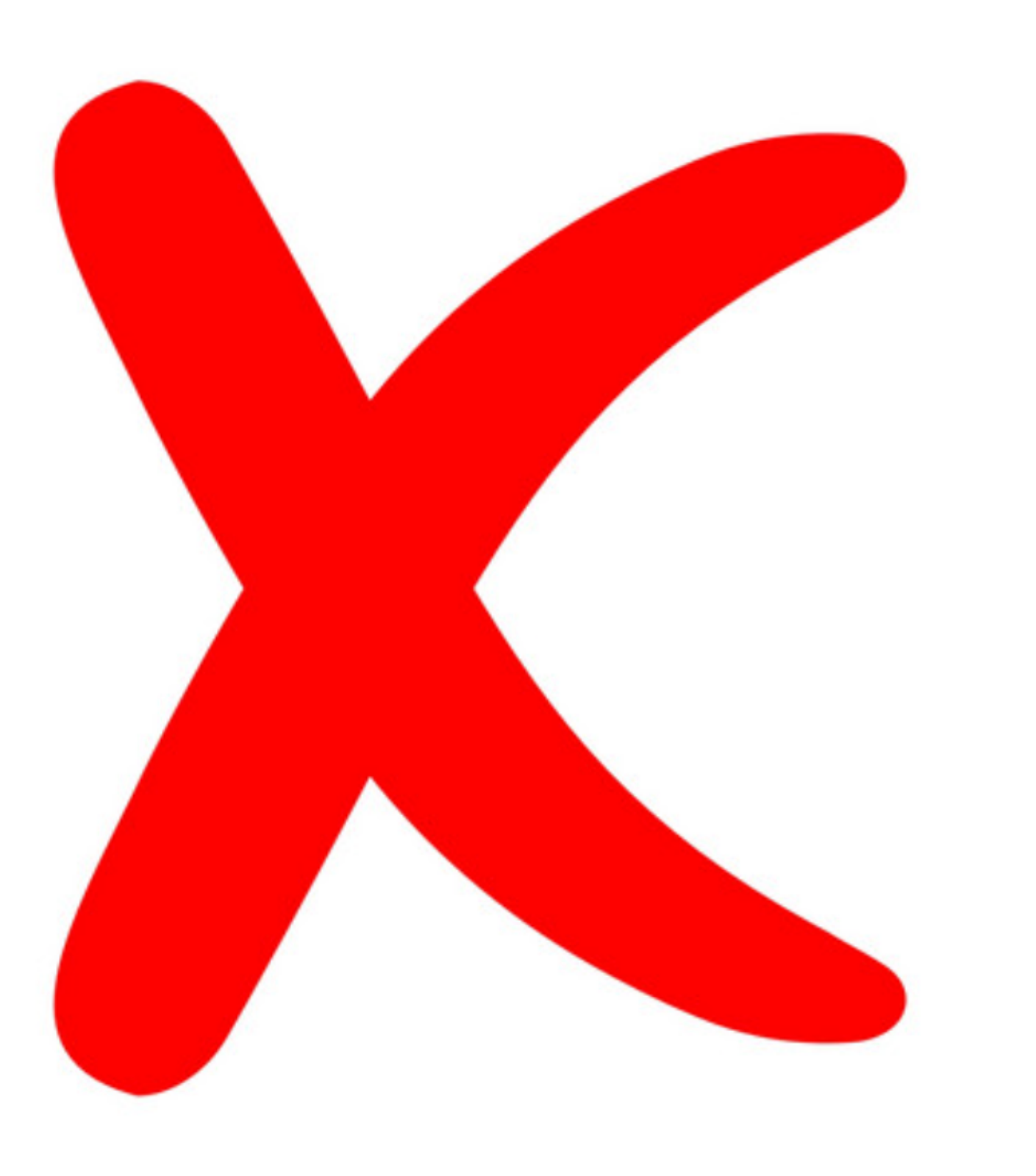}}}%

\def\system{TideCache\xspace}
\newcommand{\elasticpool}{Transparent Cache Pool\xspace}

\renewcommand{\algorithmicrequire}{\textbf{Input:}}
\renewcommand{\algorithmicensure}{\textbf{Output:}}

%% file: sections/0-abstract.tex
\begin{abstract}
Hybrid architectures have become mainstream among open-weight language models. Their global-attention and bounded-state layers maintain heterogeneous caches whose demands vary with context length and concurrency. We present \system, a serving system on CPU--GPU superchips with three components for allocating GPU memory among caches and managing their residency across GPU and host memory. The \emph{\elasticpool} redistributes physical pages among cache components behind fixed virtual addresses, preserving cache layouts and backend interfaces. The \emph{Cache Streamer} uses computation intervals provided by bounded-state layers to prefetch the next global-attention layer's prefix into GPU slots shared across global-attention layers, reducing GPU residency while overlapping layer-wise transfers with computation. Generated KV remains resident. The \emph{Request Scheduler} selects resident or streamed execution and triggers capacity adjustments according to request size and queue state. Together, they let HBM released by streaming accommodate other cache components and additional requests. Implemented in SGLang and evaluated on GB300 and GH200 across seven model/platform configurations, \system\ completes up to 2.04$\times$ as many agent turns, reduces overall P50 per-token latency by 20--59\%, and improves agent P50 TTFT by 1.40--3.05$\times$ compared with SGLang with HiCache.



\end{abstract}

%% file: sections/1-intro.tex
\section{Introduction}
\label{sec:intro}

\suppressfloats[t]
\begin{figure}[t]
\centering
\includegraphics[width=\columnwidth]{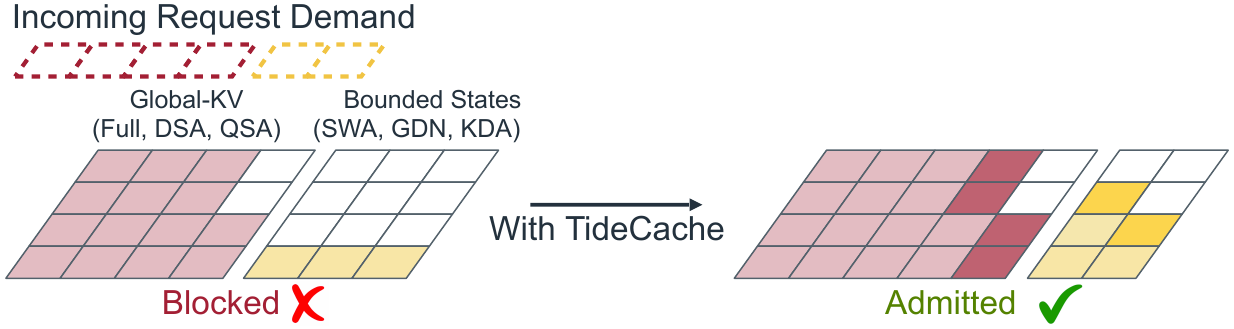}
\par\vspace{1pt}
{\footnotesize (a) Sharing capacity across cache components.\par}
\vspace{3pt}
\includegraphics[width=\columnwidth]{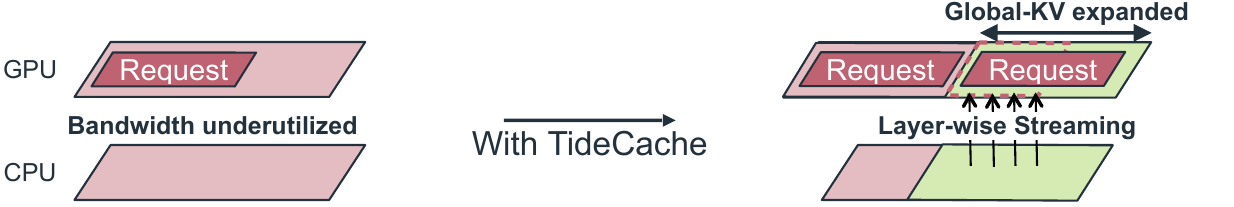}
\par\vspace{1pt}
{\footnotesize (b) Utilizing idle CPU--GPU bandwidth during decode.\par}
\caption{Two opportunities for efficient hybrid-model serving: sharing cache capacity and exploiting idle decode-time bandwidth.}
\label{fig:teaser}
\vspace{-5pt}
\end{figure}

Hybrid architectures have become mainstream among open-weight language models. Kimi K3~\cite{kimik3}, Ring-2.6~\cite{ring26}, GLM-5.3-Flash~\cite{glm53flash}, Qwen3.8~\cite{qwen3.8} and GPT-OSS~\cite{gptoss} keep a global-attention layer only once every few layers and fill the rest with sliding-window attention or recurrent layers such as Mamba, Gated DeltaNet and Kimi Delta Attention~\cite{mamba,gdn,kda}. We refer to the latter as \emph{bounded-state layers}, since their cache memory per request is bounded independently of context length. These models serve everyday chat, coding agents, document analysis and automated data labeling. Such applications differ in context reuse, interaction structure, arrival patterns and concurrency~\cite{parrot,servegen}. A coding agent accumulates tool outputs across turns, document analysis retains reference material, and chat or labeling services may process many independent requests. These behaviors create different demands on global-attention KV, window caches and recurrent states; no application label alone determines which component limits serving capacity.

CPU--GPU superchips offer a larger memory hierarchy for these workloads by coupling GPU memory with host memory through high-bandwidth interconnects. The bandwidth of this link has grown by an order of magnitude in five years (\tab{tab:link}): from 64~GB/s over PCIe~4.0 on A100 to 900~GB/s over NVLink-C2C on Grace Hopper, and 1.8~TB/s on Vera Rubin~\cite{nvidia-grace-hopper,nvidia-blackwell-ultra,nvidia-vera-rubin}. This allows GPUs to access host-resident data through both direct memory accesses and explicit transfers. Host memory thus provides an additional tier for the different cache components maintained by hybrid models.

\begin{figure}[t]
\centering
\includegraphics[width=\columnwidth]{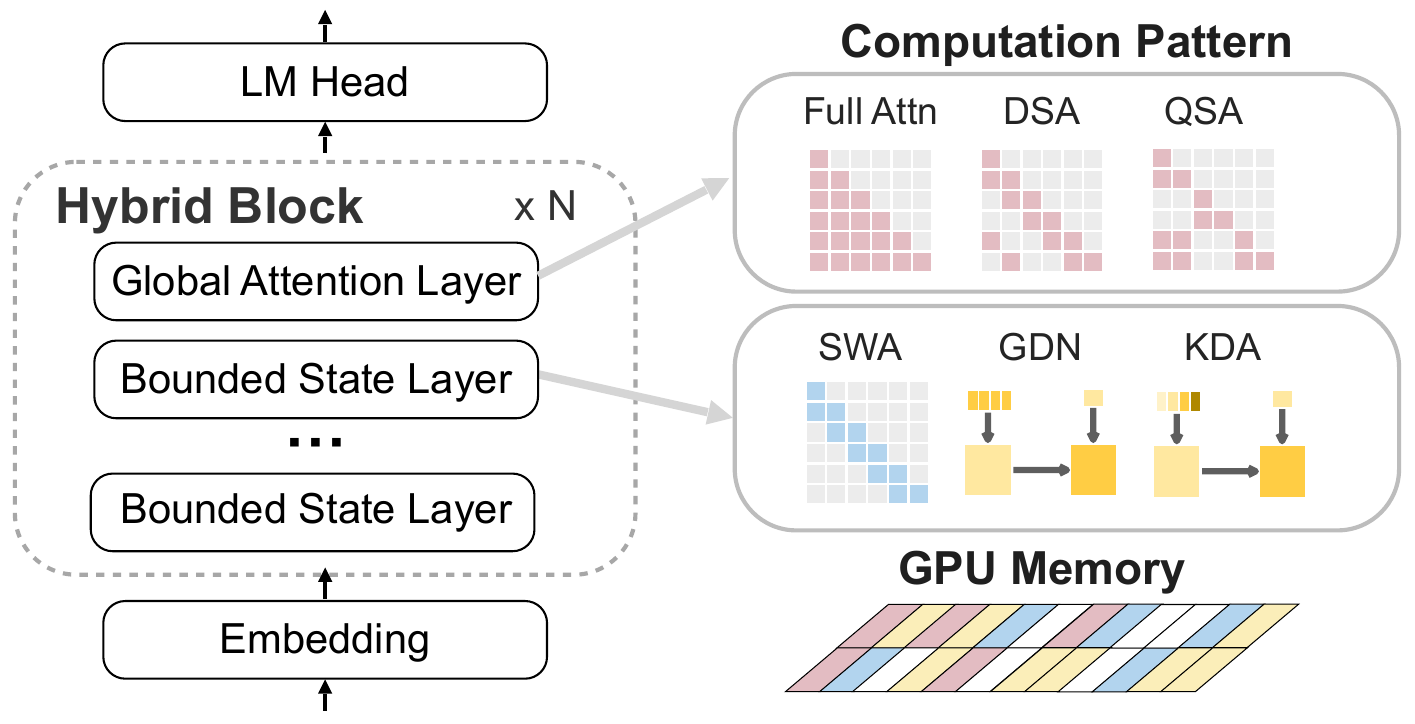}
\caption{Illustration of a typical hybrid LLM architecture. Global-attention layers (full, DSA or QSA~\cite{qsa}) hold a global-KV that grows with the context; bounded-state layers (sliding window, GDN, KDA) hold a window KV or recurrent state of fixed size per request. All cache components share GPU memory.}
\label{fig:hybrid}
\vspace{-5pt}
\end{figure}

Existing systems improve hybrid models' memory utilization through cache allocation~\cite{sglang,vllm}. Static partitioning reserves a fixed budget for each component and can leave one pool underused while another runs out of space (\fig{fig:teaser}a). Dynamic sharing schemes, such as SGLang's Unified Memory Pool and vLLM's Hybrid KV Cache Manager, reduce this imbalance~\cite{sglang-unified,vllm-hybrid}. However, sharing an arena across cache layouts can require changes to indexing and memory access, complicating integration with attention backends and~other~serving~features.

A complementary line of work uses offloading to reduce GPU memory residency. HiCache~\cite{hicache} preserves reusable contexts in a hierarchical cache and restores them for later execution, reducing repeated prefill computation. However, restored prefixes remain GPU-resident during subsequent decoding. HiSparse~\cite{hisparse} reduces active requests' GPU footprint by fetching selected KV entries during decode, but is specialized for sparse attention. General offloading systems such as FlexGen and Pie~\cite{flexgen,pie} support larger inference batches through tensor placement and transfer scheduling, but do not leverage the cache access patterns of hybrid models.

To this end, we present \textbf{\system}, a hybrid-model serving system with three components: the \textbf{\elasticpool}, a \textbf{Cache Streamer}, and a \textbf{Request Scheduler}. Our key observation is that bounded-state layers create intervals of computation during which the historical KV of the next global-attention layer is not yet needed. The Cache Streamer exploits idle CPU--GPU bandwidth during decode to prefetch host-resident prefixes into GPU slots shared across global-attention layers (\fig{fig:teaser}b). Overlapping these transfers with computation expands effective global-KV capacity: requests with reusable host-resident prefixes shrink their GPU footprint, while newly generated KV stays resident. To make the released capacity available to other cache components, the \elasticpool redistributes physical pages behind the reserved virtual addresses of the hybrid subpools. The \elasticpool transparently supports existing cache layouts and attention backend interfaces, without requiring changes to either. The two memory mechanisms work together: layer-wise KV streaming frees HBM that can hold window caches and recurrent states for additional requests, and a larger decode batch provides more computation with which to overlap subsequent transfers. The Request Scheduler coordinates these mechanisms, selecting resident or streamed execution and triggering subpool capacity adjustments according to request and queue conditions.

We implement \system\ in SGLang and evaluate seven model/platform configurations on GB300 and GH200 superchips without model-specific kernel changes. Across mixtures of Chat and Coding Agent traffic, \system\ completes up to 2.04$\times$ as many agent turns, reduces overall P50 per-token latency by 20--59\%, and improves agent P50 TTFT by 1.40--3.05$\times$ compared with SGLang with HiCache. The \elasticpool achieves 96.3--113.6\% of the turns of the better of two hand-tuned partitions at each fixed load; under a shifting workload on GH200, the full system completes 59\% more agent turns and reduces agent P50 TTFT by 78\% relative to SGLang with HiCache. At equal batch size, the system adds at most 6.3\% to decode-step latency.

This paper makes three contributions:
\begin{itemize}[leftmargin=*,itemsep=-5pt,topsep=-5pt]
\item We identify idle partition capacity and idle decode-time CPU--GPU bandwidth as two opportunities in hybrid serving.
\item We design the \elasticpool, the Cache Streamer and the Request Scheduler, and build them behind fixed virtual addresses in SGLang.
\item We evaluate seven model/platform configurations on two superchip platforms across varying workloads.
\end{itemize}

%% file: sections/2-background.tex
\section{Background}
\label{sec:background}

\begin{table}[t]
\centering
\small
\setlength{\tabcolsep}{4pt}
\begin{tabular}{@{}llrr@{}}
\toprule
Platform (year) & CPU--GPU interconnect & GB/s & vs.\ A100 \\
\midrule
A100 (2020) & PCIe 4.0 $\times$16 & 64 & 1$\times$ \\
H100 (2022) & PCIe 5.0 $\times$16 & 128 & 2$\times$ \\
GH200 (2023) & NVLink-C2C & 900 & 14$\times$ \\
GB300 (2025) & NVLink-C2C & 900 & 14$\times$ \\
VR300 (2026) & NVLink-C2C & 1,800 & 28$\times$ \\
\bottomrule
\end{tabular}
\caption{Bidirectional bandwidth of the CPU--GPU interconnect across NVIDIA generations: one PCIe $\times$16 slot per GPU, and one NVLink-C2C interface per Grace superchip (shared by the two GPUs of a GB300 superchip).}
\label{tab:link}
\end{table}

\subsection{Hybrid Model Inference}
\label{sec:bg-hybrid}

An autoregressive model runs a prefill phase, which builds per-layer state for the input, and a decode phase, which produces one token per request per step and updates that state. What the state is depends on the layer (\fig{fig:hybrid}). A \emph{global-attention} layer lets every token attend to all earlier positions, so its Key--Value (KV) cache grows with the context. A \emph{bounded-state} layer keeps a state whose size is independent of the context~\cite{abc}: a sliding-window layer keeps the KV of the most recent window, and a recurrent layer such as Mamba, Gated DeltaNet or Kimi Delta Attention keeps a fixed-size state updated in place. The model stacks the hybrid block of \fig{fig:hybrid} repeatedly, placing one global-attention layer every few bounded-state layers. We call the data each layer type keeps a \emph{cache component}: global-KV, window KV and recurrent state; where the distinction does not matter we write \emph{state} for the two bounded components. For Ring-2.6-1T at TP4 with BF16 caches, a 67K-token context uses about 0.72~GiB of global-KV per GPU, while one recurrent-state slot occupies 35~MiB. The evaluated SGLang configuration budgets five such slots per running request for admission (175~MiB, or 0.17~GiB), independent of context length.

\subsection{Hybrid Cache Management}
\label{sec:bg-cache}

Serving systems reuse the cached state of a prefix to avoid prefilling it again. In a hybrid model a global-KV hit alone is not enough: a bounded-state layer can resume only from a checkpoint at the matched position, so every component must be consistent at the same boundary. SGLang's unified radix cache coordinates matching, insertion and eviction of all components under one prefix tree~\cite{sglang,unifiedradixcache}, and its memory pool owns the tensors and decides how much GPU memory each component receives, either through a partition fixed at start-up or through a unified pool whose boundary moves at run time~\cite{sglang-unified}. HiCache extends the tree to host memory: evicted prefixes and checkpoints are backed up there and loaded back before a session's next turn runs~\cite{hicache}. \system\ replaces only the memory-pool layer and inherits the radix cache and HiCache.

\subsection{Offloading on CPU--GPU Superchips}
\label{sec:bg-offload}

On Grace superchips the GPU reaches host memory over NVLink-C2C; one GB300 GPU reads pinned host memory at 211~GB/s, and four GPUs reading concurrently reach 182--192~GB/s each. Two ways of using this path differ in what stays on the GPU. \emph{Request-level restore} copies a session's prefix back once before the turn runs; the prefix is then resident for the whole turn~\cite{hicache}. \emph{Layer-wise loading} keeps on the GPU only the layers about to execute and transfers the rest again in every decode step~\cite{pie}; it is hidden only if each layer's copy finishes before that layer starts, which depends on how much compute separates the layers that need the data.

%% file: sections/3-motivation.tex
\section{Motivation}
\label{sec:motivation}

\begin{figure}[t]
\centering
\includegraphics[width=\columnwidth]{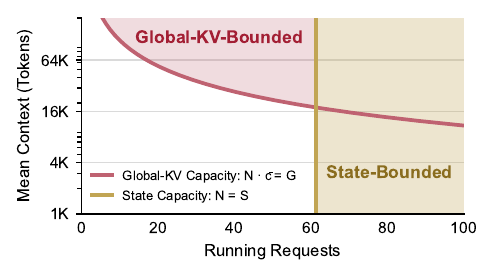}
\vspace{-10pt}
\caption{Loads that one GPU of Ring-2.6-1T (TP4) can run under SGLang's default partition (state : global-KV = 0.9). The state pool caps the request count at $S = 61$; the global-KV pool caps the product of request count and context length at $G = 1.09$M tokens.}
\label{fig:motivation}

\end{figure}

\subsection{Cache Demand Depends on the Workload}
\label{sec:mot-demand}

Different workloads stress different parts of a hybrid model's cache. A busy chat or data-labeling service may have many requests active at once, each requiring its own window KV or recurrent state. Coding agents or document analysis may instead retain substantial context, increasing the global-KV footprint per request. These examples describe demand patterns, not fixed application categories. For $N$ running requests with mean context $\bar c$, global-KV demand grows with $N\bar c$, whereas bounded-state demand grows only with $N$. A fixed partition can therefore exhaust one component while leaving another underused. \fig{fig:motivation} illustrates this imbalance using Ring-2.6-1T under SGLang's default partition. The state pool limits concurrency to 61 requests even when most of global-KV remains unused. Conversely, at 64K tokens per request, global-KV holds only 17 requests while most of the state pool remains unused. A different partition moves these limits but retains the tradeoff: giving more memory to one component leaves less for the others. In a mixed workload, demand shifts with each request source's arrival rate, so a partition tuned for one mix can become a bottleneck for another. Repartitioning cannot increase the GPU's total cache budget: when combined demand exceeds HBM capacity, it only changes which requests wait.

\begin{table*}[!t]
\centering
\small
\begin{tabular}{@{}lccc@{}}
\toprule
 & \makecell{SGLang\\{\scriptsize\cite{sglang}}} & \makecell{vLLM\\{\scriptsize\cite{vllm}}} & \system \\
\midrule
Technique & \makecell{Unified Memory Pool\\{\scriptsize\cite{sglang-unified-blog}}} & \makecell{Hybrid KV Manager\\{\scriptsize\cite{vllm-hybrid}}} & \elasticpool \\
Dynamic Sharing & \cmark & \cmark & \cmark \\
Copy-Free Repartitioning & \xmark & \cmark & \cmark \\
Transparency & \xmark & \xmark & \cmark \\
Streaming Compatibility & \xmark & \xmark & \cmark \\
\bottomrule
\end{tabular}
\caption{Comparison of hybrid cache management solutions.}
\label{tab:pools}
\vspace{-15pt}
\end{table*}

\subsection{The CPU--GPU Link Is Idle During Decode}
\label{sec:mot-host}

HiCache uses host memory between turns: a session's prefix is restored in full before its next turn is admitted and stays resident until the turn ends. During decode, which dominates the turn, the host-to-device link carries almost nothing. Its capacity is large: a GB300 GPU reads host memory at 211~GB/s, so one 57~ms decode step of Ring-2.6-1T at the loads of \sect{sec:eval} could move about 12~GB per GPU at that bandwidth. A hybrid model can use this capacity because only its global-attention layers read the prefix. Between two of them, 7 bounded-state layers on Ring-2.6 compute without needing any global-KV, so the next layer's prefix can be transferred in that interval and the previous layer's can be discarded. A request served this way needs device memory for a few global-attention layers at a time rather than for all of them. More computation per decode step allows more transfer to overlap. Diverse workloads can create this opportunity: concurrent requests enlarge the decode batch while other requests retain reusable host-resident prefixes. The relevant properties are batch computation and cache demand, rather than application identity.

%% file: sections/4-method.tex
\section{Design}
\label{sec:method}

Motivated by the observations in \sect{sec:motivation}, we build \system, a serving system with three components. The \emph{\elasticpool} (TCP, \sect{sec:method-layout}) repartitions GPU memory at run time by moving physical pages to the cache component that blocks admission. The \emph{Cache Streamer} (\sect{sec:method-streaming}) serves streamed requests by copying each global-attention layer's prefix into a shared slot just before that layer runs. The \emph{Request Scheduler} (\sect{sec:method-admission}) decides whether the first request in SGLang's waiting queue (the \emph{head request}) runs with its prefix resident on the GPU or streamed from host memory, and which cache component blocks it if neither fits. All three operate on one memory layout whose defining property is that every cache component keeps fixed virtual addresses: attention backends, the page table, CUDA graphs and HiCache address the cache as they always have, and \system\ only changes which physical pages back an address and which host rows are present at an address during a step.

\begin{figure*}[t]
\centering
\includegraphics[width=0.92\textwidth]{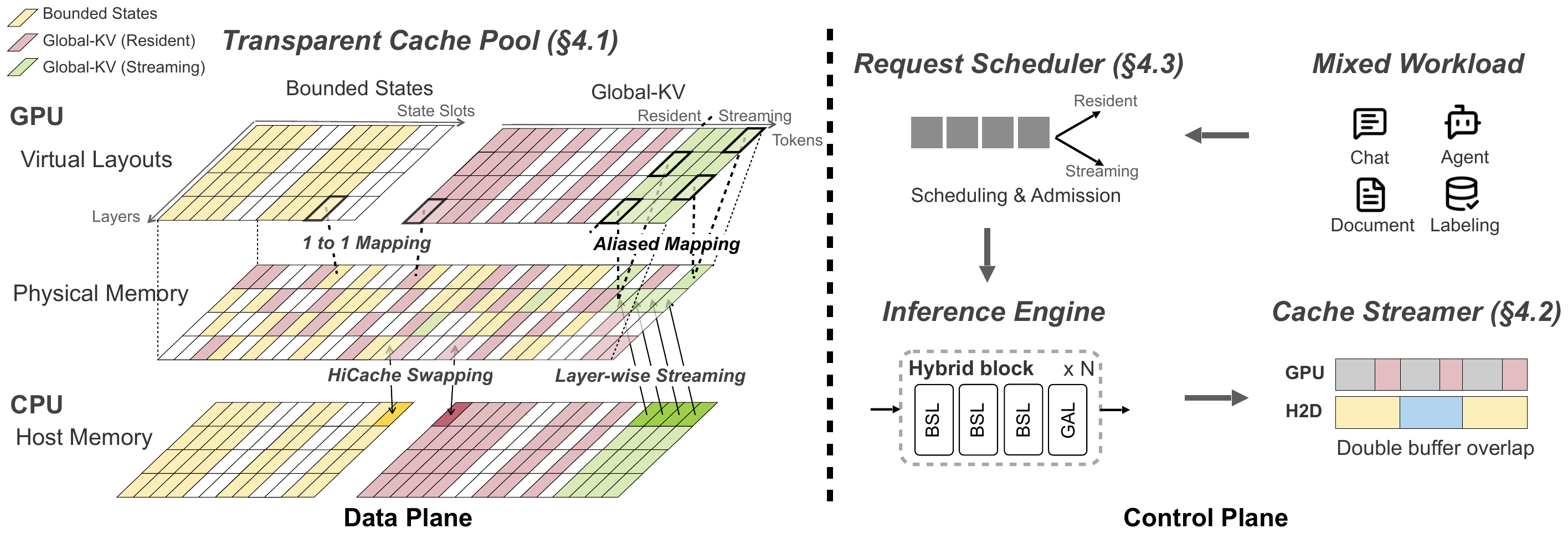}
\caption{\system overview: the Request Scheduler controls resident- and streaming-region admission; the \elasticpool manages cache backing, and the Cache Streamer feeds shared GPU slots from host memory.}
\label{fig:overview}
\vspace{-10pt}
\end{figure*}

\fig{fig:overview} presents \system's architecture. The Request Scheduler controls admission to the resident and streaming regions, shown by its two labeled arrows, and dispatches admitted requests to the inference engine. The data plane exposes virtual cache layouts, physical GPU memory and host cache. The \elasticpool manages a shared arena of GPU pages, repartitioning it among cache components as demand changes. For the head request, the scheduler selects resident or streamed execution, or reports the blocking cache component to the \elasticpool for growth between forward passes. Before each forward pass the Cache Streamer assigns slot rows to the batch's streamed prefixes and prepares their copies, while HiCache restores resident prefixes and backs up finished turns as before.

\subsection{Transparent Cache Pool}
\label{sec:method-layout}
\label{sec:method-elastic}

As \sect{sec:mot-demand} shows, a fixed partition can block requests even when another cache component has unused capacity. The \elasticpool addresses this imbalance by dynamically repartitioning a shared GPU physical page arena among global-KV, window KV and recurrent state. Each cache component has a \emph{subpool}: a fixed virtual address range backed by a variable number of physical pages. To \emph{grow} a subpool, the \elasticpool maps additional arena pages into its virtual range; to \emph{shrink} one, it unmaps pages that no active request uses, making them available to another subpool. This adjusts capacity without relocating live data or changing~its~virtual~addresses.

The physical arena is allocated once in 2~MiB pages through the CUDA virtual memory API~\cite{cuda-vmm}. Each subpool reserves more virtual address space than any physical share it will receive, and its tensors span this reservation. The allocator uses only rows whose backing pages are mapped. Rows retain SGLang's index addressing, so forward passes are independent of physical page locations.

\paragraph{Transparent dynamic sharing.} \tab{tab:pools} contrasts how dynamic sharing affects existing cache interfaces.\footnote{Audited at SGLang revision 25ce8063 (the evaluation baseline) and vLLM revision 94a29d3. SGLang's Unified Memory Pool is disabled by default and requires \texttt{--enable-unified-memory}.} SGLang's \emph{Unified Memory Pool} (UMP) allocates global-KV and bounded state from opposite ends of a raw \texttt{uint8} buffer. It packs all layers' KV for a page into one \emph{page envelope}; reclaiming fragmented space can require copying live envelopes and updating address mappings. This page-major layout requires stride-compatible attention backends and logical-to-physical index translation~\cite{sglang-unified,sglang-unified-blog}. In this revision, UMP rejects HiCache and LMCache: host pools are not initialized, and host offloading lacks logical-to-physical slot translation~\cite{sglang-unified-compat}. vLLM's \emph{Hybrid KV Cache Manager} shares uniformly sized pages among cache groups without relocating live blocks~\cite{vllm-hybrid,vllm-agentx}. KV-transfer connectors must implement \texttt{SupportsHMA} to handle per-group block lists and~their~asynchronous~release~\cite{vllm-hma-interface}.

These designs expose pool-specific layouts or grouping to cache consumers. \textbf{TCP confines repartitioning to physical page mappings.} It preserves per-layer tensors, virtual addresses and cache indexing, enabling copy-free repartitioning without pool-specific adaptation of attention backends or HiCache. The same virtual layout natively supports aliased mappings for layer-wise streaming (\sect{sec:method-streaming}), preserving existing cache interfaces.

\begin{figure}[t]
\centering
\includegraphics[width=\columnwidth]{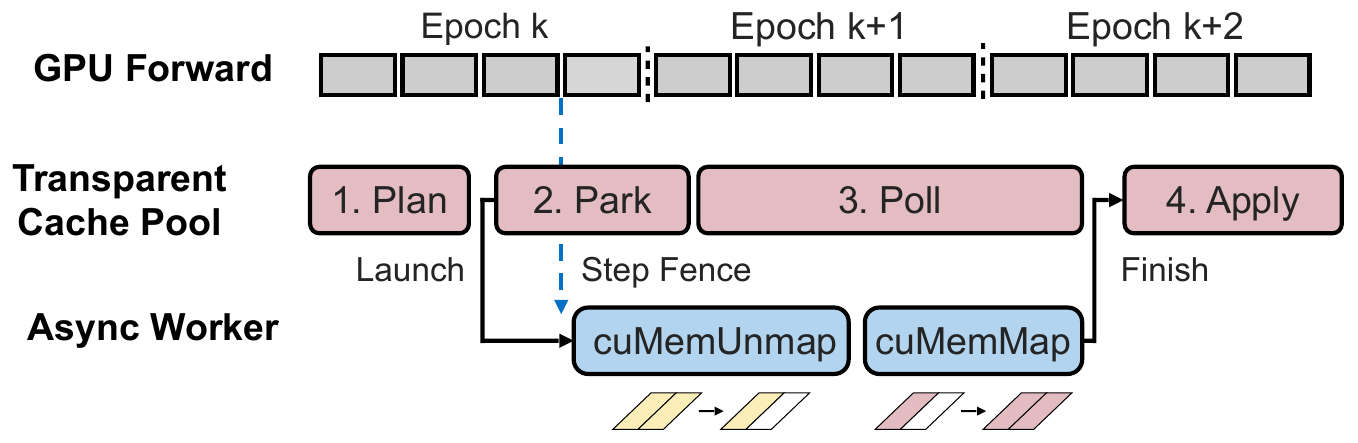}
\caption{One page move across epochs: plan and park on the scheduler thread, unmap and map on a worker after the step fence, apply at the next epoch in which every rank is idle.}
\label{fig:pagemove}
\end{figure}

\paragraph{Move protocol.} The \elasticpool rebalances capacity once per \emph{epoch} of $E = 8$ forward passes, with at most one move in flight (\fig{fig:pagemove}). It \emph{plans} the receiver, the amount and the donors, and \emph{parks} the donor rows by removing them from their allocator's free list; rows in use are never touched. A \emph{step fence} waits until every forward pass issued before the park has completed. A worker thread then calls \texttt{cuMemUnmap} on the parked pages and \texttt{cuMemMap} to back the receiver's lowest parked rows, followed by \texttt{cuMemSetAccess}; these driver calls do not wait for the GPU and touch only memory no batch may use, so forward passes continue. The pool manager \emph{polls} the worker at each epoch and \emph{applies} the move, un-parking the receiver's rows, in the first epoch in which every rank's worker is done. Each epoch starts with one all-reduce of a busy flag, so all ranks apply the same move before the same forward pass and build identical batches.

\paragraph{Policy.} The \elasticpool adopts a \emph{failure-driven} repartitioning policy to direct capacity to the component blocking admission while limiting cache eviction. When the Request Scheduler cannot admit the head request because a subpool lacks capacity, it asks the \elasticpool to grow that subpool. This subpool is the \emph{receiver}; pages come from the following sources, in order of increasing reclamation cost:
\begin{enumerate}[label=(\arabic*),leftmargin=*,itemsep=-5pt,topsep=-5pt]
\item \textbf{Free pages}, which are not assigned to any subpool.
\item \textbf{Idle capacity in other subpools}, reclaimed by unmapping pages from unused rows. These subpools act as \emph{donors} and shrink as the receiver grows.
\item \textbf{Cold cached prefixes in donor subpools}, defined as prefixes untouched for $\tau$ seconds. Evicting a cold prefix and its checkpoints releases their cache rows.
\end{enumerate}
The \elasticpool exhausts each source before proceeding to the next. If growth can be funded without eviction, its size is scaled by the waiting-queue length (capped at $k$) to avoid admitting only one request per multi-epoch move. If eviction is required, growth is limited to the head request's needs to avoid discarding more cached data than necessary. Hot cached prefixes are not reclaimed as donors, because discarding a prefix reused after a tool gap can force recomputation. A subpool shrinks only to supply a receiver. Before admission fails, TCP can also grow a subpool whose free space runs low by one small step from idle capacity or cold cache in other subpools, sparing hot prefixes from later eviction.

\subsection{Cache Streamer}
\label{sec:method-streaming}

Streaming every request's context at every decode step can exceed the link's transfer budget. For example, the GPT-OSS-120B ablation (\sect{sec:eval-streaming-overhead}) sustains a batch of 29 requests with 8K-token contexts on GH200. Their BF16 global-KV occupies 8.76~GB across 18 global-attention layers. Transferring this entire batch at 400~GB/s would take 21.9~ms per step, exceeding the measured 19.7~ms decode step with selective streaming. Layer-wise pipelining alone therefore cannot hide all-batch transfers at this operating point. The Cache Streamer instead streams only a subset of requests selected by the Request Scheduler, keeping the rest resident. This raises three questions: how to reduce only these requests' physical footprint, how to move their changing KV addresses without breaking the decode graph, and how much transfer can be hidden.

\begin{figure}[t]
\centering
\includegraphics[width=\columnwidth]{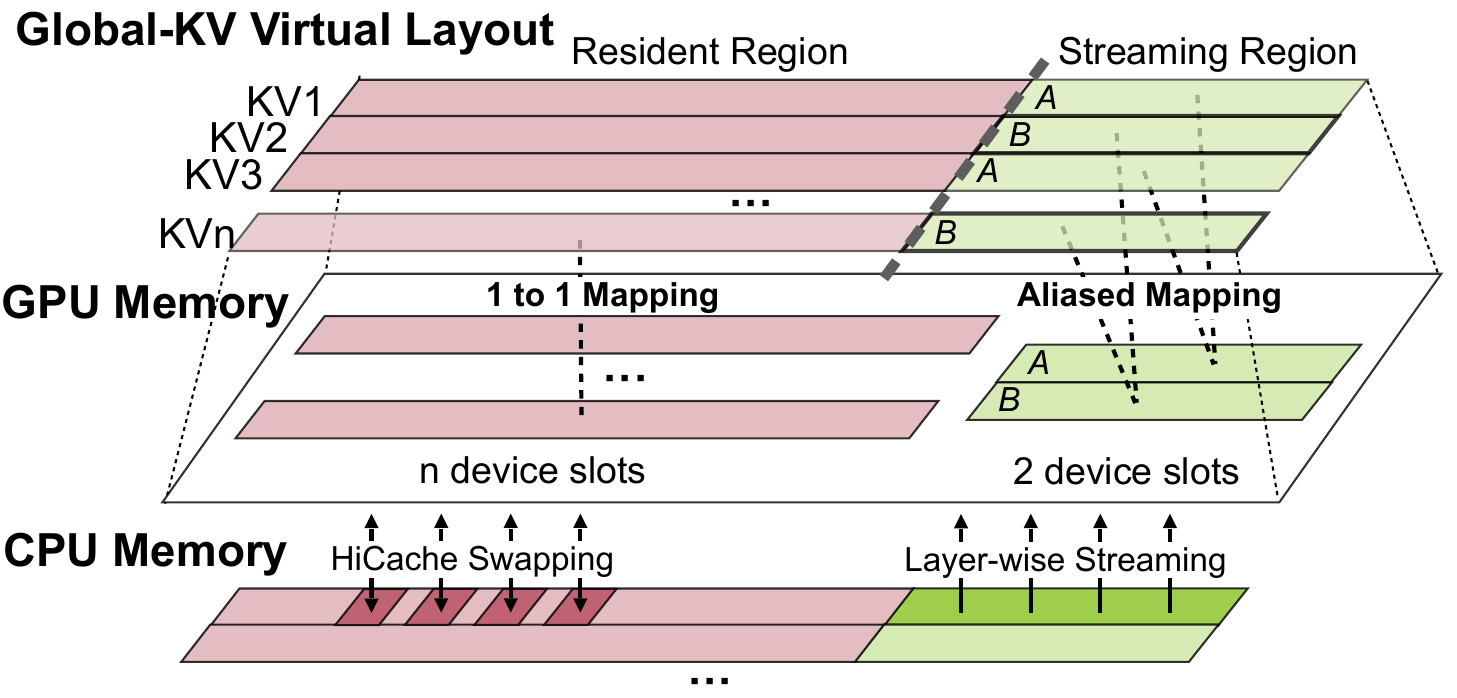}
\caption{The global-KV subpool (KV1, KV2, etc. denote successive layers' caches): a resident region with its own pages per layer and a streaming region whose $L$ layer ranges alias $M=2$ slots; HiCache restores into the former, streaming feeds the latter.}
\label{fig:layout}
\vspace{-2pt}
\end{figure}

\paragraph{How can selected requests use less GPU memory?}
The global-KV subpool separates \emph{resident} and \emph{streaming} regions while retaining the same per-layer tensor interfaces (\fig{fig:layout}). The resident region has independent backing pages for every layer, managed by TCP. It holds the prefixes of requests served without streaming and the newly generated KV of all requests. Only historical prefixes already available in host memory are streamed; generated KV is backed up by HiCache at the end of the turn.

At start-up, the streaming region's virtual ranges for $L$ global-attention layers are mapped onto $M=2$ physical slots, with layer $\ell$ using slot $(\ell-1) \bmod M$. These \emph{aliased mappings} let different layers address the same physical pages through their original per-layer tensors. A streamed prefix occupies only $M/L$ of its full-layer global-KV footprint on the GPU. Its rows hold one layer's prefix at a time; generated tokens remain in separate resident rows. The slot capacity is set at start-up using the bound in \eqn{eq:bound}, and the alias mappings remain fixed during execution. TCP can repartition the resident backing without changing these mappings.

\begin{figure}[t]
\centering
\includegraphics[width=\columnwidth]{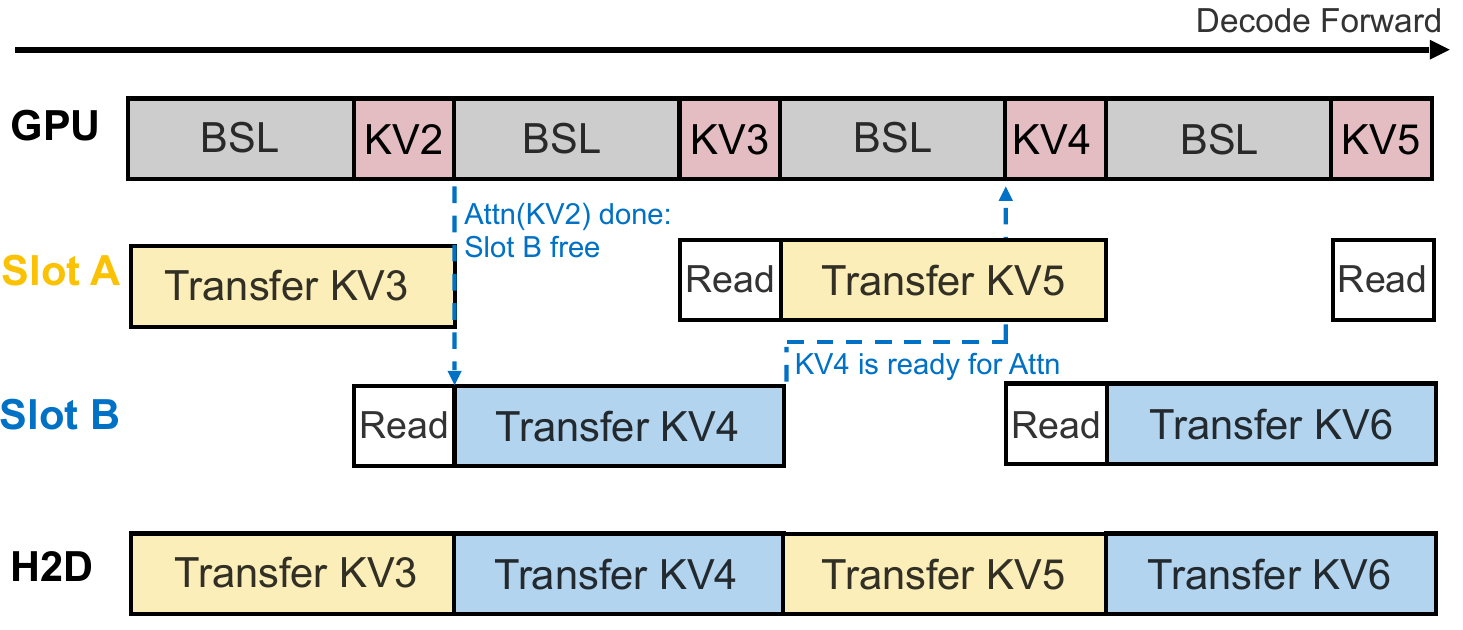}
\caption{Double-buffered overlap with two slots: reading KV2 releases slot B for KV4's transfer, which must finish before attention uses KV4. Bounded-state layers (BSL) and global attention provide computation to overlap with H2D.}
\label{fig:pipeline}
\end{figure}

These slots support the double-buffered overlap pipeline in \fig{fig:pipeline}. A single H2D stream fills one slot while the GPU reads the other, using both bounded-state computation and global attention to hide transfers. For example, after attention finishes reading KV2 from slot B, the copy stream can overwrite it with KV4 while computation proceeds through KV3 and the intervening bounded-state layers. Realizing this schedule requires coordinating slot reuse and KV availability without interrupting the decode graph.

\begin{figure}[t]
\centering
\includegraphics[width=\columnwidth]{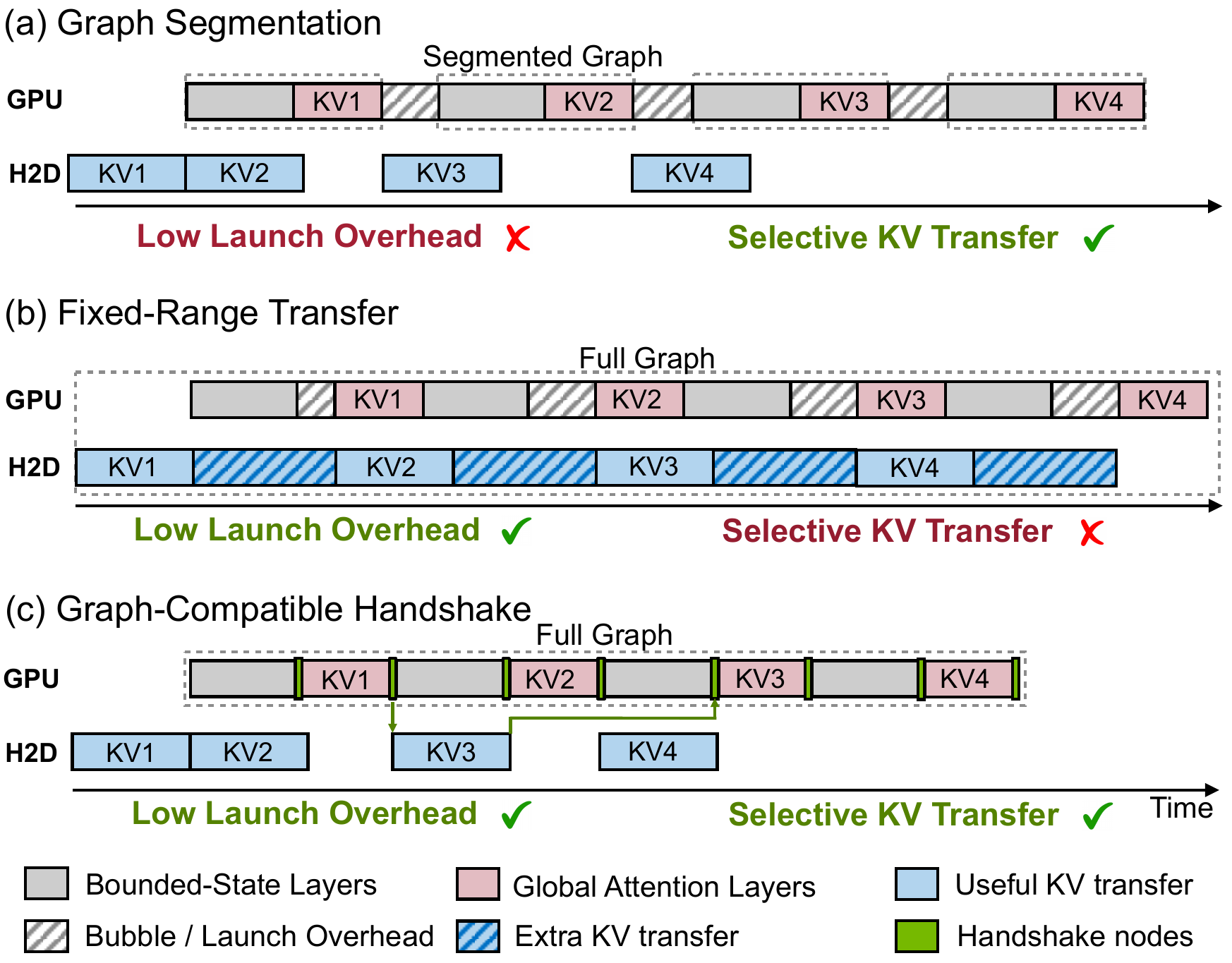}
\caption{Selective KV transfer with low launch overhead: (a) Graph Segmentation transfers selected KV but incurs repeated launches; (b) Fixed-Range Transfer retains low launch overhead but moves extra KV; (c) Graph-Compatible Handshake achieves both. Timelines are schematic.}
\label{fig:handshake}
\end{figure}

\paragraph{How can selective KV transfer retain low launch overhead?}
The pipeline must transfer only the batch's streamed prefixes, whose host addresses, destination rows and lengths can change each step. A CUDA graph, however, captures a fixed set of operations and dependencies. The difficulty is bidirectional: a transfer must wait until the previous reader releases its slot, and a later attention operation inside the same decode step must wait for that transfer. \fig{fig:handshake} compares three designs for selective KV transfer and low launch overhead.

\emph{Graph Segmentation} (\fig{fig:handshake}a) divides the decode graph into segments and submits dynamic copies and their dependencies between segment launches. The boundaries let the host submit a copy and its completion record before enqueueing the consumer's wait. This supports selective KV transfer, but repeated graph launches and host coordination can expose submission bubbles. \emph{Fixed-Range Transfer} (\fig{fig:handshake}b) captures copies of a fixed host range into the slots alongside computation, retaining low launch overhead through a full decode graph. The copies also move inactive rows and unused capacity, however, wasting link bandwidth and delaying attention when the extra transfers exceed the available overlap.

Our \emph{Graph-Compatible Handshake} (\fig{fig:handshake}c) retains both properties: low launch overhead and selective KV transfer. Before replay, the Cache Streamer assigns the batch's streamed prefixes to contiguous slot rows, updates their page-table entries, and coalesces adjacent host rows into source--destination copy runs. This preparation changes metadata only; it adds no KV staging copy. The Cache Streamer submits these runs with \texttt{cudaMemcpyBatchAsync}~\cite{cuda-memcpy-batch} on the H2D stream, copying directly from the host cache into the slots without transferring KV for requests outside the batch.

Two persistent flags coordinate each layer's use of a slot. On the copy stream, a transfer waits for and clears \textsc{Free} from the previous reader, copies the selected rows, and publishes \textsc{Ready}. Inside the graph, small kernels wait for and clear \textsc{Ready} before the layer's first KV access and publish \textsc{Free} after its final read. The arrows in \fig{fig:handshake}c illustrate both dependencies: completing KV1's reads permits KV3's transfer, whose completion permits KV3's reads. Each flag is cleared when consumed, preventing a previous step's signal from satisfying the current wait. The first use of a slot waits for its last reader in the previous step; both slots start free.

The host submits the first transfer into each slot, launches the full decode graph once, and then submits the remaining transfers. Execution-time flag checks let the graph wait for these transfers even though their submission follows its launch. Thus DMA addresses and lengths can vary while graph nodes and flag addresses stay fixed. The host need not wait for individual layers, and a late transfer stalls its consumer until the KV is ready. Each auxiliary kernel uses one warp and replays within the graph, without a separate host launch. This preserves selective-transfer efficiency without fragmenting the decode graph; the combined implementation overhead is evaluated in \sect{sec:eval-streaming-overhead}.

\paragraph{How much can streaming expand effective global-KV capacity?}
\label{sec:method-capacity}
For a fixed GPU cache budget, the useful expansion is bounded by how much host-resident KV can be delivered within each decode step. With $L$ global-attention layers, $b$ bytes per row and layer, computation time $T$ and sustained link bandwidth $\mathrm{BW}$, streaming $A$ rows transfers $L A b$ bytes per step. To hide these transfers, we require $L A b \le \eta\,\mathrm{BW}\,T$, where $\eta < 1$ leaves scheduling slack. The resulting \emph{hidden bound} $C$ limits admission (\sect{sec:method-admission}). With $M=2$ streaming slots, the bound and relative capacity gain are
\begin{equation}
\label{eq:bound}
C \;=\; \frac{\eta\,\mathrm{BW}\,T}{L\,b},
\qquad
r \;=\; \frac{C\,(1 - M/L)}{D},
\end{equation}
where $D$ is the capacity, in rows, of an entirely resident global-KV pool with the same GPU budget. Slots for $C$ streamed rows consume the equivalent of $MC/L$ resident rows, leaving a net gain of $C(1-M/L)$. Ignoring page rounding, effective capacity therefore reaches $(1+r)D$, provided the slots fit within the GPU budget. This is an aggregate bandwidth bound; transfers must also meet each layer's deadline.

In the GPT-OSS-120B ablation (\sect{sec:eval-streaming-overhead}), $L=18$, $b=2$~KiB, $M=2$, and $D=147{,}648$. The sizing calibration ($T=13.16$~ms, $\mathrm{BW}=400$~GB/s, $\eta=0.9$) gives $C\approx128.5$K and \textbf{77\% capacity gain}. The configured 126,976 streaming rows plus 133,056 resident rows yield 260,032 tokens after page rounding---\textbf{1.76$\times$ the original capacity}.

For a fixed model and slot count, more bandwidth or longer computation increases the expansion; a larger per-token KV footprint or original resident capacity reduces the relative gain. We size $C$ using an offline decode calibration at the expected batch. Extra capacity serves host-resident historical prefixes; newly generated KV remains resident. Higher concurrency follows when global-KV capacity limits admission and the other cache components have room.

\begin{algorithm}[t]
\caption{Request Admission}
\label{alg:admit}
\begin{algorithmic}[1]
\REQUIRE $r{:}\,(n_s, n_f, h)$; queue $Q$; streamed rows $A$; bound $C$
\STATE \textbf{if} $n_s > \mathrm{free}(S)$ \textbf{then} \textsc{Grow}($S$); \textbf{return}
\IF{$h \ge h_{\min}$ \textbf{and} $\mathrm{room}(G) - h < \textsc{Pressure}(Q)$}
  \IF{$A + h \le C$}
    \STATE \textsc{Admit}($r$, streamed); $A \gets A + h$; \textbf{return}
  \ELSIF{$h \le C$}
    \STATE \textsc{Skip}($r$); \textbf{return}
  \ENDIF
\ENDIF
\IF{$n_f + h \le \mathrm{room}(G)$}
  \STATE \textsc{Admit}($r$, resident)
\ELSE
  \STATE \textsc{Grow}($G$)
\ENDIF
\STATE \textsc{Pressure}($Q$) $= \sum_{q \in Q} \mathrm{need}(q) + N\bar\ell/2 + \lambda\,\bar\ell\,\bar p\,\phi$
\STATE \textbf{procedure} \textsc{Grow}($X$) \COMMENT{TCP, next epoch}
  \STATE \hspace{1em} $n \gets \mathrm{need}(X) - \mathrm{free}(X)$
  \STATE \hspace{1em} \textbf{if} $n \le \mathrm{idle}$ \textbf{then} $n \gets n \cdot \min(|Q|, k)$
  \STATE \hspace{1em} \textsc{Take}($n$) \COMMENT{free, idle, cold}
  \STATE \hspace{1em} \textsc{Park}; \textsc{Fence}; \textsc{Remap}; \textsc{Apply}
\STATE \textbf{end procedure}
\end{algorithmic}

\end{algorithm}

\subsection{Request Scheduler}
\label{sec:method-admission}

The Request Scheduler coordinates TCP and the Cache Streamer to admit more requests within the available GPU capacity and transfer budget. For the head request $r$, it considers its state slots $n_s$, uncached global-KV rows $n_f$, and host-cached prefix length $h$. A \emph{resident} turn holds $n_f+h$ resident rows. A \emph{streamed} turn keeps $n_f$ rows resident and uses $h$ streaming-slot rows, transferring $h\,b$ bytes per layer each step; total streamed rows must stay within $C$ (\sect{sec:method-capacity}). If capacity is insufficient, TCP can grow the blocking subpool using free pages, idle capacity, or cold cache. \alg{alg:admit} gives the decision order.

A request is streamed only when the resident region is \emph{contended}: \textsc{Pressure} projects the region's demand over one turn from the rows the waiting requests need, the decode growth of the $N$ running requests over a mean output length $\bar\ell$, and the arrivals during that turn ($\lambda$ per step, prompt length $\bar p$, cache-miss fraction $\phi$), all count-based running averages of events every rank observes. A request selected for streaming skips this iteration if the slots are full, allowing requests needing other resources to proceed. $\mathrm{room}(G)$ counts the resident region's free rows and cached rows with a host copy, minus a reserve for in-flight prefills; $h_{\min}$ is the minimum~prefix~size~worth~streaming.

%% file: sections/5-evaluation.tex
\section{Evaluation}
\label{sec:eval}

\subsection{Experimental Setup}
\label{sec:eval-setup}

\paragraph{Implementation.}\label{sec:method-impl} \system\ adds about 9K lines to SGLang, mainly in a memory-pool package. It supports multi-head, latent and sparse-indexed global attention, sliding-window caches and recurrent states, while preserving attention backends and the CUDA graph runner.

\paragraph{Testbed.} We use two superchip platforms. The GB300 node has four NVIDIA GB300 GPUs with 288~GB of HBM3e each, connected by NVLink, and two 72-core Grace CPUs with 960~GB of LPDDR5X host memory; each GPU reaches host memory over NVLink-C2C at a measured 211~GB/s host-to-device. The GH200 superchip pairs one 72-core Grace CPU with 480~GB of LPDDR5X and one H100 GPU with 96~GB of HBM3 over NVLink-C2C.

\begin{table}[t]
\centering
\small
\setlength{\tabcolsep}{4pt}
\begin{tabular}{@{}llcl@{}}
\toprule
Model & Hybrid combination & TP & Platform \\
\midrule
Kimi K3 (48L) & MLA : KDA = 1 : 3 & 4 & GB300 \\
Ring-2.6-1T & MLA : LA = 1 : 7 & 4 & GB300 \\
GLM-5.3-Flash & DSA : KDA = 11 : 34 & 1 & GB300 \\
Qwen3.8-2.4T (36L) & GQA : GDN = 1 : 3 & 4 & GB300 \\
GLM-5.3-Flash & DSA : KDA = 11 : 34 & 4 & GH200 \\
Qwen3.8-27B & GQA : GDN = 1 : 3 & 1 & GH200 \\
GPT-OSS-120B & GQA : SWA = 1 : 1 & 1 & GH200 \\
\bottomrule
\end{tabular}
\caption{Hybrid layer combinations and hardware configurations of the evaluated models. MLA: multi-head latent attention; DSA: DeepSeek sparse attention; GQA: grouped-query attention; KDA: Kimi Delta Attention; GDN: Gated DeltaNet; LA: Lightning linear attention; SWA: sliding-window attention.}
\label{tab:setup}
\end{table}

\paragraph{Models.} \tab{tab:setup} lists seven model/platform configurations, covering latent, grouped-query and sparse-indexed global attention over three kinds of recurrent layer and sliding-window attention. Kimi K3 and Qwen3.8-2.4T use their first 48 and 36 layers to fit in HBM, preserving the cache geometry per layer. SGLang's memory fraction is left at its normal deployment value (0.80--0.95).

\paragraph{Workloads and metrics.} We construct synthetic mixtures from ShareGPT request traces~\cite{sharegpt} and multi-turn Claude Code traces released by SemiAnalysis~\cite{claude-code-traces}. The former have 896--1152 input tokens and 703 output tokens on average, and arrive as an open-loop Poisson process. The latter have 40K--110K-token contexts and 8--28 turns; closed-loop replay preserves tool gaps and feeds outputs into subsequent turns. Both sources share one first-come-first-served queue, without application labels or priorities supplied to the scheduler. We vary the independent-request arrival rate and the number of concurrent sessions; configuration labels use \emph{req/s $\times$ sessions}. After warm-up (360~s on GB300; at least 600~s on GH200), the measurement window is 600~s, or 900~s for the shifting workload. \emph{Turns/s} is the number of agent-session turns completed within the window divided by the window duration. \emph{P50 per-token latency} is the median, over all requests submitted within the window, of end-to-end request latency divided by actual output-token count. \emph{P50 TTFT} is the median submission-to-first-token latency for session requests submitted within that window. Turns/s and TTFT thus describe agent sessions, while per-token latency covers the full mixture; both latencies include prefill and queueing, and they count the submitted cohort's completions after the window.

\subsection{End-to-End Performance}
\label{sec:eval-e2e}

We evaluate performance under memory pressure across seven configurations, then examine adaptation to a workload whose demand changes during execution.

\begin{figure*}[!t]
\centering
\includegraphics[width=\textwidth]{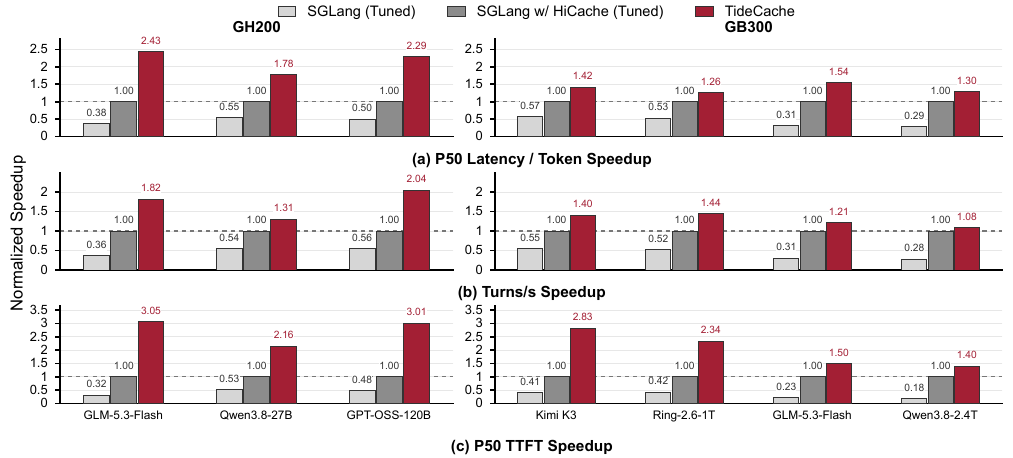}
\caption{End-to-end performance across seven model/platform configurations. (a) P50 latency / token speedup across all requests; (b) turns/s and (c) P50 TTFT speedups for agent sessions. Both baselines use their best tested split ratios. Values are normalized to SGLang w/ HiCache (Tuned). All panels are higher-is-better. GH200 is on the left, GB300 on the right.}
\label{fig:exp1}
\vspace{-15pt}
\end{figure*}

\paragraph{Performance under memory pressure.} \fig{fig:exp1} uses one load point per configuration, where session prefixes exceed the resident global-KV capacity. Without HiCache, repeated prefix computation occupies the GPU and delays subsequent requests. HiCache avoids much of this recomputation, but each restored prefix remains in GPU memory throughout its turn, limiting the sessions that can proceed concurrently. \system\ relieves this capacity constraint by reusing GPU memory across global-attention layers. Relative to SGLang w/ HiCache (Tuned), it completes up to 2.04$\times$ as many turns, reduces overall P50 per-token latency by 20--59\%, and improves agent P50 TTFT by 1.40--3.05$\times$. \system\ reduces TTFT by relieving memory pressure and shortening the wait to serve session prefixes. Turn gains reach up to 2.04$\times$ on GH200 and 1.44$\times$ on GB300, consistent with the larger capacity benefit when less HBM remains for global-KV. \sect{sec:eval-split} examines the effect of baseline partition tuning.

\begin{figure*}[!t]
\centering
\includegraphics[width=\textwidth]{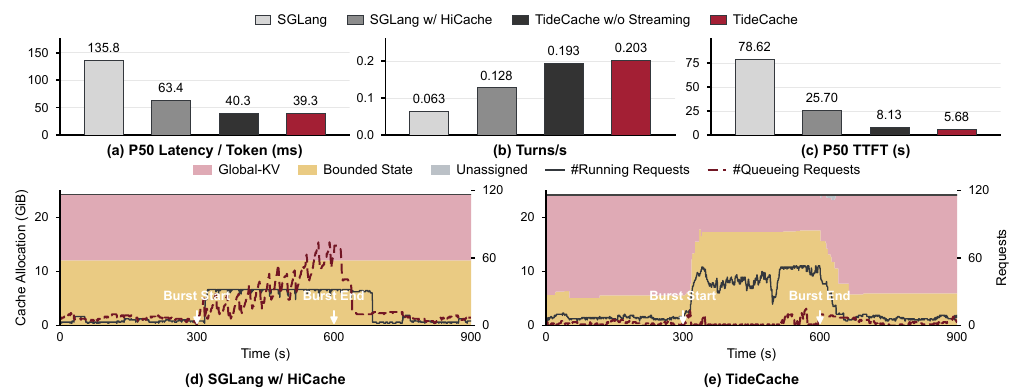}
\caption{Qwen3.8-27B on GH200 under a shifting workload, with a burst at 300--600~s. (a) Overall P50 latency / token; (b) turns/s and (c) P50 TTFT for agent sessions. (d--e) Physical GPU capacity allocated to each cache component, including occupied and free space within its pool, alongside running and queueing request counts. Unassigned denotes pages outside both pools.}
\label{fig:exp3}
\vspace{-10pt}
\end{figure*}

\paragraph{Adaptation to workload changes.}\label{sec:eval-shifting} \fig{fig:exp3} runs Qwen3.8-27B on GH200 after a 900~s warm-up. The 900~s measurement comprises three 300~s phases: independent-request arrivals rise from 0.05 to 1.6 req/s and then return to 0.05 req/s. Ten agent workers replay ten sessions before and after the burst; new agent turns pause during the burst while in-flight requests finish. This shifts demand between global-KV and bounded state. The static baselines retain a state : global-KV ratio of 1.0, while both \system\ variants start at 0.6 and adapt online. SGLang, SGLang w/ HiCache, \system{} w/o streaming, and \system\ complete 57, 115, 174, and 183 agent turns, respectively. Relative to HiCache, \system\ completes 59\% more turns, reduces overall P50 per-token latency from 63.4 to 39.3~ms (38\%), and reduces agent P50 TTFT from 25.70 to 5.68~s (78\%).

The timelines expose the difference between fixed and adaptive allocation. HiCache's pool capacities remain unchanged as the workload shifts. TCP grows \system's bounded-state allocation from about 5.6 to 17.6~GiB during the burst, then returns it to about 6.0~GiB as agent activity resumes. Roughly half of the GPU cache budget changes owners, making capacity available to the component under pressure. The static baselines reach their native 32-request running limit; HiCache's burst queue peaks at 74 requests, compared with 15 for \system. Streaming further expands effective global-KV capacity for session prefixes: relative to the no-streaming variant, it increases completed turns from 174 to 183 and reduces agent P50 TTFT from 8.13 to 5.68~s (30\%).

\subsection{Ablation Study}
\label{sec:eval-ablation}

We examine the \elasticpool, sensitivity to the split ratio, and the implementation overhead in ablations.

\begin{figure}[t]
\centering
\includegraphics[width=\columnwidth]{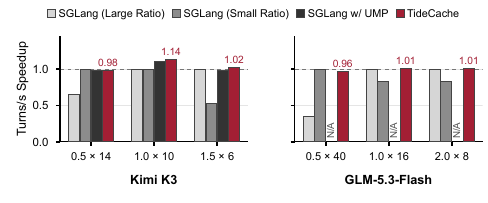}
\vspace{-20pt}
\caption{Dynamic cache partitioning on Kimi K3 and GLM-5.3-Flash. Turns/s is normalized to the better fixed partition; configurations denote req/s $\times$ concurrent sessions.}
\label{fig:exp2}

\end{figure}

\paragraph{Standalone TCP Performance.}\label{sec:eval-partition} \fig{fig:exp2} isolates TCP by disabling host caching and layer-wise streaming, comparing it with fixed partitions and UMP. Both TCP and UMP dynamically share GPU capacity; we test whether TCP matches UMP's throughput while preserving existing cache interfaces. UMP's HiCache integration was incomplete in the evaluated revision, so we include it only in this GPU-only comparison. Each model uses three mixtures, with large/small fixed state : global-KV ratios of approximately 1.3/0.26 for Kimi K3 and 2.8/0.85 for GLM-5.3-Flash. TCP starts from one allocation per model and adapts online. Across six points, it achieves 96.3--113.6\% of the better fixed partition's turns/s without per-mixture tuning. It matches UMP on Kimi K3, with 0--3.7\% higher turns/s. N/A denotes GLM's DSA backend, unsupported by UMP revision \texttt{25ce8063}; TCP supports its existing layout.

\begin{figure}[!t]
\centering
\includegraphics[width=\columnwidth]{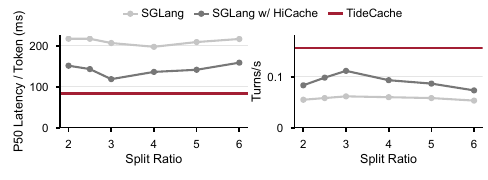}
\vspace{-10pt}
\caption{Sensitivity to the state : global-KV split ratio on Kimi K3. Horizontal lines show \system\ starting from ratio 3.}
\label{fig:exp1b}

\end{figure}

\paragraph{Sensitivity to Split Ratio.}\label{sec:eval-split} \fig{fig:exp1b} illustrates the baseline tuning used in \fig{fig:exp1}, sweeping the split ratio from 2 to 6 on Kimi K3. SGLang w/ HiCache peaks at ratio 3 with 67 completed turns, the setting used in the main comparison, versus 44--59 at the other tested ratios: allocating too little memory to state limits admission, while allocating too little to global-KV limits resident prefixes. The \system\ reference completes 94 turns, 40\% above even the best tested baseline, and has 29.5\% lower overall P50 per-token latency at that starting ratio. Thus tuning the baseline's fixed partition does not close the gap.

\begin{figure}[!t]
\centering
\includegraphics[width=\columnwidth]{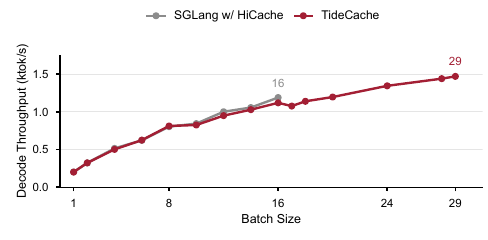}
\caption{Median decode throughput versus batch size on GPT-OSS-120B / GH200 with 8K-token contexts.}
\label{fig:ablation}
\vspace{-5pt}
\end{figure}

\paragraph{System Overhead.}\label{sec:eval-streaming-overhead} \fig{fig:ablation} compares median decode throughput over three runs per batch size, with the same GPU cache budget and SWA : global-KV split. The measurement includes the combined implementation on this path: transfer submission, page-table updates, and CUDA-graph coordination. At batch sizes both systems sustain, \system's decode-step time differs from SGLang w/ HiCache by $-$1.1\% to $+$6.3\%. The added capacity compensates for this cost: effective global-KV capacity rises from 148K to 260K tokens, and the largest sustained batch grows from 16 to 29. Peak decode throughput rises from 1.19K to 1.47K tokens/s, a 24\% gain.

%% file: sections/6-related.tex
\section{Related Work}
\label{sec:related}

\paragraph{Hybrid Cache Management.}
Static partitioning fixes each cache component's GPU budget at start-up. SGLang's Unified Memory Pool instead shares a byte buffer among subpools and compacts live slots by copying~\cite{sglang-unified}; vLLM's Hybrid KV Cache Manager dynamically allocates shared blocks to cache groups~\cite{vllm-hybrid}. Jenga uses LCM-sized parent pages in a two-level allocator for heterogeneous caches~\cite{jenga}. These designs enable sharing through software-level cache layouts and mappings. vAttention first backed a contiguous virtual KV cache with physical pages on demand through the CUDA virtual memory API~\cite{vattention}, and Prism remaps physical pages between models sharing a GPU~\cite{prism}. \system applies the same primitive across the cache components of one hybrid model, adds a region whose pages are shared by all global-attention layers, and drives the remapping from the admission queue.


\paragraph{KV Cache Offloading.}
FlexGen schedules weights, activations and KV across GPU, CPU and disk~\cite{flexgen}; ZeRO-Inference streams layer weights from CPU or NVMe to support large inference batches~\cite{zeroinference}. Pie swaps the KV of transformer layers between host and device so that the active set exceeds GPU memory~\cite{pie}; InfiniGen and HiSparse fetch only the KV that attention will actually read~\cite{infinigen,hisparse}; SuperInfer rotates requests through the GPU and uses both transfer directions to keep latency bounded on superchips~\cite{superinfer}. \system\ exploits hybrid structure: only global-attention layers hold context-proportional data, while bounded-state computation provides time to overlap its transfer. This makes per-step streaming affordable when decode provides sufficient computation.

%% file: sections/7-conclusion.tex
\section{Conclusion}
\label{sec:conclusion}

We presented \system: the \elasticpool transparently repartitions HBM, the Cache Streamer overlaps layer-wise transfers with computation, and the Request Scheduler coordinates admission. Across seven model/platform configurations on GB300 and GH200, it completes up to 2.04$\times$ as many agent turns, reduces overall P50 per-token latency by 20--59\%, and improves agent P50 TTFT by 1.40--3.05$\times$ over SGLang with HiCache.

%% file: sections/8-ai-disclosure.tex
\section*{AI Usage Disclosure}
We used OpenAI Codex interactively to assist with literature discovery and source checking, manuscript drafting and revision, terminology and presentation choices, and figure design. Codex also assisted in writing and modifying plotting and data-analysis scripts, inspecting recorded experimental results, and checking consistency between figures and reported metrics. This assistance was iterative and guided by the authors, covering multiple manuscript sections and evaluation figures. AI assistance did not extend to system implementation, experiment design or execution, or workload construction. The authors retain full responsibility for the research claims, experimental methodology, reported results, citations, and final manuscript.

%% file: refs.bib
@misc{mamba,
      title={Mamba: Linear-Time Sequence Modeling with Selective State Spaces}, 
      author={Albert Gu and Tri Dao},
      year={2024},
      eprint={2312.00752},
      archivePrefix={arXiv},
      primaryClass={cs.LG},
      url={https://arxiv.org/abs/2312.00752}, 
}

@misc{gdn,
      title={Gated Delta Networks: Improving {Mamba2} with Delta Rule}, 
      author={Songlin Yang and Jan Kautz and Ali Hatamizadeh},
      year={2025},
      eprint={2412.06464},
      archivePrefix={arXiv},
      primaryClass={cs.CL},
      url={https://arxiv.org/abs/2412.06464}, 
}

@misc{kda,
      title={{Kimi Linear}: An Expressive, Efficient Attention Architecture}, 
      author={{Kimi Team}},
      year={2025},
      eprint={2510.26692},
      archivePrefix={arXiv},
      primaryClass={cs.CL},
      url={https://arxiv.org/abs/2510.26692}, 
}

@misc{kimik3,
      title={{Kimi K3}: Open Frontier Intelligence}, 
      author={{Kimi Team}},
      year={2026},
      eprint={2607.24653},
      archivePrefix={arXiv},
      primaryClass={cs.CL},
      url={https://arxiv.org/abs/2607.24653}, 
}

@misc{ring26,
      title={Ling and Ring 2.6 Technical Report: Efficient and Instant Agentic Intelligence at Trillion-Parameter Scale}, 
      author={Ang Li and Ben Liu and Bin Han and Bin Hu and Bin Jing and Binbin Hu and Bing Li and Cai Chen and Caizhi Tang and Changxin Tian and Chao Huang and Chao Zhang and Chen Liang and Chen Qian and Chengfu Tang and Chengyao Wen and Chilin Fu and Chunwei Wu and Cong Zhang and Cunyin Peng and Daixin Wang and Dalong Zhang and Deng Zhao and Dingnan Jin and Dingyuan Zhu and Donghao Zhang and Fan Yuan and Fangzheng Zhao and Fanzhuang Meng and Feifan Wu and Feng Xu and Fengbin Fang and Gangshan Wang and Guodong Yang and Hailin Zhao and Haitao Wang and Haitao Zhang and Hanxiao Zhang and Hanzi Wang and Hao Dai and Hao Liu and Hao Qian and Hao Wu and Haoxiong Liu and Haoyu Xu and Heng Zhang and Hong Liu and Hongliang Zhang and Hongrui Liu and Hongxun Li and Hongzhi Ruan and Huaidong Xiong and Huihuang Zheng and Huikang Tang and Jia Guo and Jia Li and Jia Liu and Jiameng Wang and Jiaming Liu and Jiannan Shi and Jianping Wei and Jiaolong Yang and Jiapeng Wang and Jie Gao and Jie Wang and Jiewei Wu and Jin Yang and Jinjin Li and Jinjing Huang and Jinquan Sun and Jinyao Chen and Juanhui Tu and Jun Liu and Jun Mei and Jun Xu and Jun Zhou and Junjie Ou and Junnan Sipan and Junpeng Fang and Kaihong Zhang and Kaiqin Hu and Ke Shi and Kuan Xu and Kun Tang and Kunlong Chen and Lanyin Mei and Lei Chen and Lei Liang and Lei Xu and Li Tang and Liang Jiang and Liangcheng Fu and Lihui Zhang and Linfeng Shi and Lintao Ma and Liyuan Liu and Longfei Li and Longfei Zheng and Lu Liu and Lu Yu and Man Li and Meiqi Zhu and Meng Li and Mengjie Gao and Mengshu Sun and Mingming Yin and Mingyang Zhang and Mingyuan Fan and Nuo Xu and Pan Tang and Peijie Jiang and Peilong Zhao and Peng Lin and Pingping Liu and Qi Zuo and Qian Zhao and Qiang Cheng and Qianggang Cao and Qiaoben Bao and Qing Cui and Qingyuan Yang and Qitao Shi and Qiyin Huang and Qizheng Zhou and Quan Wan and Runyuan Zhao and Shaomian Zheng and Shaowei Wei and Shengnan Zhang and Shuaicheng Li and Shujie Li and Shuo Zhang and Sikang Bian and Tianchu Yao and Tiange Xu and Tianshu Wang and Ting Guo and Tinghao Wang and Tingwei Huang and Tong Zhao and Tongkai Yang and Wang Hong and Wanli Gu and Wei Lu and Weichang Wu and Weiguang Han and Weiquan Li and Wenbo Shen and Wenjing Fang and Wenzhi Tang and Xiang Shu and Xiao Shi and Xiaodong Yan and Xiaolu Zhang and Xiaopei Wan and Xiaqing Sun and Xin Zhao and Xingyu Lu and Xinxing Yang and Xinyao Tang and Xinyu Kong and Xinyu Liu and Xiong Xu and Xuan Sun and Xudong Han and Xudong Wang and Xujie Shen and Yalin Zhang and Yangyang Hou and Yankun Ren and Yao Zhao and Ye Chen and Yeyang Chen and Yibo Cao and Yifan Zuo and Yijie Chen and Ying Li and Yingjie Song and Yingxue Li and Yiqi Wang and Yixuan Sun and Yizhu Xiao and Yongfei Xu and Yu Liu and Yuchen Fang and Yue Gao and Yue Yu and Yue Zhang and Yuqi Zhang and Yuxiao He and Yuxiao Lu and Yuxin Tian and Yuxuan Li and Yuzhuo Fu and Zhankai Xu and Zhaoxin Huan and Zhenduo Zhang and Zhengke Gui and Zhengyu Huang and Zhenjun Ma and Zhenxuan Pan and Zheping Qu and Zhibo Zhu and Zhidong Fan and Zhigang Huangfu and Zhihao Wang and Zhiqiang Zhang and Zhizhen Liu and Zhuyan Zhou and Zibin Lin and Zihang Zeng and Zihao Wang and Zilong Wang and Ziqi Liu and Zitao Xuan and Zixuan Cheng and Zujie Wen and Zuoli Tang},
      year={2026},
      eprint={2606.15079},
      archivePrefix={arXiv},
      primaryClass={cs.CL},
      url={https://arxiv.org/abs/2606.15079}, 
}

@misc{qwen3.8,
    title  = {{Qwen3.8-Max}: A New Bar for Coding and Cowork},
    author = {{Qwen Team}},
    year   = {2026},
    month  = {August},
    url    = {https://qwen.ai/blog?id=qwen3.8}
}

@misc{gptoss,
      title={gpt-oss-120b \& gpt-oss-20b Model Card},
      author={OpenAI},
      year={2025},
      eprint={2508.10925},
      archivePrefix={arXiv},
      primaryClass={cs.CL},
      url={https://arxiv.org/abs/2508.10925}, 
}

@misc{glm53flash,
  author       = {{Z.ai}},
  title        = {{GLM-5.3-Flash}: Frontier Intelligence, Flash Cost},
  year         = {2026},
  howpublished = {\url{https://z.ai/blog/glm-5.3-flash}},
  note         = {Accessed: 2026-10-08}
}

@misc{pie,
      title={Pie: Pooling {CPU} Memory for {LLM} Inference}, 
      author={Yi Xu and Ziming Mao and Xiangxi Mo and Shu Liu and Ion Stoica},
      year={2024},
      eprint={2411.09317},
      archivePrefix={arXiv},
      primaryClass={cs.LG},
      url={https://arxiv.org/abs/2411.09317}, 
}

@misc{hisparse,
      title={{HiSparse}: Scaling Sparse-Attention Decoding with Hierarchical {KV} Cache Management}, 
      author={Zhiqiang Xie and Zhangheng Huang and Tingwei Huang and Ziyi Xu and Ruiyang Ma and Christos Kozyrakis},
      year={2026},
      eprint={2608.07009},
      archivePrefix={arXiv},
      primaryClass={cs.DC},
      url={https://arxiv.org/abs/2608.07009}, 
}

@misc{jenga,
      title={Jenga: Effective Memory Management for Serving {LLM} with Heterogeneity}, 
      author={Chen Zhang and Kuntai Du and Shu Liu and Woosuk Kwon and Xiangxi Mo and Yufeng Wang and Xiaoxuan Liu and Kaichao You and Zhuohan Li and Mingsheng Long and Jidong Zhai and Joseph Gonzalez and Ion Stoica},
      year={2025},
      eprint={2503.18292},
      archivePrefix={arXiv},
      primaryClass={cs.DC},
      url={https://arxiv.org/abs/2503.18292}, 
}

@misc{superinfer,
      title={{SuperInfer}: {SLO}-Aware Rotary Scheduling and Memory Management for {LLM} Inference on Superchips}, 
      author={Jiahuan Yu and Mingtao Hu and Zichao Lin and Minjia Zhang},
      year={2026},
      eprint={2601.20309},
      archivePrefix={arXiv},
      primaryClass={cs.DC},
      url={https://arxiv.org/abs/2601.20309}, 
}

@misc{sglang,
      title={SGLang: Efficient Execution of Structured Language Model Programs}, 
      author={Lianmin Zheng and Liangsheng Yin and Zhiqiang Xie and Chuyue Sun and Jeff Huang and Cody Hao Yu and Shiyi Cao and Christos Kozyrakis and Ion Stoica and Joseph E. Gonzalez and Clark Barrett and Ying Sheng},
      year={2024},
      eprint={2312.07104},
      archivePrefix={arXiv},
      primaryClass={cs.AI},
      url={https://arxiv.org/abs/2312.07104}, 
}

@misc{vllm,
      title={Efficient Memory Management for Large Language Model Serving with PagedAttention}, 
      author={Woosuk Kwon and Zhuohan Li and Siyuan Zhuang and Ying Sheng and Lianmin Zheng and Cody Hao Yu and Joseph E. Gonzalez and Hao Zhang and Ion Stoica},
      year={2023},
      eprint={2309.06180},
      archivePrefix={arXiv},
      primaryClass={cs.LG},
      url={https://arxiv.org/abs/2309.06180}, 
}

@misc{flexgen,
      title={FlexGen: High-Throughput Generative Inference of Large Language Models with a Single GPU}, 
      author={Ying Sheng and Lianmin Zheng and Binhang Yuan and Zhuohan Li and Max Ryabinin and Daniel Y. Fu and Zhiqiang Xie and Beidi Chen and Clark Barrett and Joseph E. Gonzalez and Percy Liang and Christopher Ré and Ion Stoica and Ce Zhang},
      year={2023},
      eprint={2303.06865},
      archivePrefix={arXiv},
      primaryClass={cs.LG},
      url={https://arxiv.org/abs/2303.06865}, 
}

@misc{zeroinference,
  author = {{DeepSpeed Team}},
  title = {{ZeRO-Inference}: Democratizing Massive Model Inference},
  year = {2022},
  month = sep,
  howpublished = {DeepSpeed Blog},
  url = {https://www.deepspeed.ai/2022/09/09/zero-inference.html}
}

@misc{infinigen,
      title={{InfiniGen}: Efficient Generative Inference of Large Language Models with Dynamic {KV} Cache Management}, 
      author={Wonbeom Lee and Jungi Lee and Junghwan Seo and Jaewoong Sim},
      year={2024},
      eprint={2406.19707},
      archivePrefix={arXiv},
      primaryClass={cs.LG},
      url={https://arxiv.org/abs/2406.19707}, 
}

@misc{unifiedradixcache,
  author       = {Huang, Zhangheng and Bao, Ke and Zhang, Yi
                  and Ouyang, Jialin and Pan, Sicheng},
  title        = {{Unified Radix Cache}: One Tree for Hybrid Model
                  Prefix Caching},
  year         = {2026},
  month        = aug,
  howpublished = {LMSYS Org Blog.
                  \url{https://www.lmsys.org/blog/2026-08-11-unified-radix-cache/}},
  note         = {Accessed: 2026-10-08}
}

@misc{sglang-unified,
  author       = {{SGLang Team}},
  title        = {Unified memory pool for hybrid models},
  year         = {2026},
  howpublished = {SGLang GitHub repository, revision 25ce8063},
  url          = {https://github.com/sgl-project/sglang/blob/25ce8063f7dbe444c4e24c562f31f86268e05a23/python/sglang/srt/mem_cache/unified_memory_pool.py}
}

@inproceedings{vattention,
  author    = {Ramya Prabhu and Ajay Nayak and Jayashree Mohan and Ramachandran Ramjee and Ashish Panwar},
  title     = {{vAttention}: Dynamic Memory Management for Serving {LLMs} without {PagedAttention}},
  booktitle = {Proceedings of the 30th ACM International Conference on Architectural Support for Programming Languages and Operating Systems (ASPLOS)},
  year      = {2025}
}

@misc{prism,
  author       = {Shan Yu and Jiarong Xing and Yifan Qiao and Mingcong Han and Yangmin Li and Yihao Zhao and Yang Zhou and Ion Stoica and Ravi Netravali},
  title        = {Prism: Unleashing {GPU} Sharing for Cost-Efficient Multi-{LLM} Serving},
  year         = {2025},
  eprint       = {2505.04021},
  archivePrefix = {arXiv},
  primaryClass = {cs.DC}
}

@inproceedings{abc,
  author    = {Hao Peng and Jungo Kasai and Nikolaos Pappas and Dani Yogatama and Zhaofeng Wu and Lingpeng Kong and Roy Schwartz and Noah A. Smith},
  title     = {{ABC}: Attention with Bounded-memory Control},
  booktitle = {Proceedings of the 60th Annual Meeting of the Association for Computational Linguistics (ACL)},
  year      = {2022}
}

@misc{nvidia-grace-hopper,
  author       = {Jonathon Evans and Michael Andersch and Vikram Sethi
                  and Gonzalo Brito and Vishal Mehta},
  title        = {{NVIDIA Grace Hopper Superchip Architecture In-Depth}},
  year         = {2022},
  month        = nov,
  howpublished = {NVIDIA Technical Blog},
  url          = {https://developer.nvidia.com/blog/nvidia-grace-hopper-superchip-architecture-in-depth/},
  note         = {Published November 10, 2022. Accessed October 24, 2026}
}

@misc{nvidia-blackwell-ultra,
  author       = {Kyle Aubrey and Nick Stam},
  title        = {{Inside NVIDIA Blackwell Ultra: The Chip Powering the AI Factory Era}},
  year         = {2025},
  month        = aug,
  howpublished = {NVIDIA Technical Blog},
  url          = {https://developer.nvidia.com/blog/inside-nvidia-blackwell-ultra-the-chip-powering-the-ai-factory-era/},
  note         = {Published August 22, 2025. Accessed October 24, 2026}
}

@misc{nvidia-vera-rubin,
  author       = {Kyle Aubrey},
  title        = {{Inside the NVIDIA Vera Rubin Platform: Six New Chips, One AI Supercomputer}},
  year         = {2026},
  month        = jan,
  howpublished = {NVIDIA Technical Blog},
  url          = {https://developer.nvidia.com/blog/inside-the-nvidia-rubin-platform-six-new-chips-one-ai-supercomputer/},
  note         = {Published January 5, 2026. Accessed October 24, 2026}
}

@misc{vllm-hybrid,
  author = {{vLLM Team}},
  title = {Hybrid {KV} Cache Manager},
  year = {2026},
  howpublished = {vLLM Design Documentation},
  url = {https://docs.vllm.ai/en/latest/design/hybrid_kv_cache_manager/},
  note = {Implementation audited at revision 94a29d3; accessed 2026-10-25}
}

@misc{vllm-hma-interface,
  author = {{vLLM Contributors}},
  title = {{SupportsHMA}: Hybrid Memory Allocator Connector Interface},
  year = {2026},
  howpublished = {vLLM source code, revision 94a29d3},
  url = {https://github.com/vllm-project/vllm/blob/94a29d316e8c1d6dcf6ff75645c2e61c7e3c83fd/vllm/distributed/kv_transfer/kv_connector/v1/base.py},
  note = {Accessed 2026-10-25}
}

@inproceedings{parrot,
  author = {Chaofan Lin and Zhenhua Han and Chengruidong Zhang and Yuqing Yang and Fan Yang and Chen Chen and Lili Qiu},
  title = {Parrot: Efficient Serving of {LLM}-based Applications with Semantic Variable},
  booktitle = {18th USENIX Symposium on Operating Systems Design and Implementation (OSDI 24)},
  year = {2024},
  pages = {929--945},
  url = {https://www.usenix.org/conference/osdi24/presentation/lin-chaofan}
}

@inproceedings{servegen,
  author = {Yuxing Xiang and Xue Li and Kun Qian and Yan Zhang and Wenyuan Yu and Ennan Zhai and Xin Jin and Jingren Zhou},
  title = {{ServeGen}: Workload Characterization and Generation of Large Language Model Serving in Production},
  booktitle = {23rd USENIX Symposium on Networked Systems Design and Implementation (NSDI 26)},
  year = {2026},
  pages = {1845--1859},
  url = {https://www.usenix.org/conference/nsdi26/presentation/xiang-servegen}
}

@misc{vllm-agentx,
  author = {{vLLM Team and Inferact}},
  title = {{vLLM} x {AgentX}: Optimizing for Real-World Agentic Serving},
  year = {2026},
  month = sep,
  howpublished = {vLLM Blog},
  url = {https://vllm.ai/blog/2026-09-08-vllm-agentx},
  note = {Accessed 2026-10-26}
}

@misc{sglang-unified-compat,
  author = {{SGLang Contributors}},
  title = {Unified Memory Pool Compatibility Checks},
  year = {2026},
  howpublished = {SGLang source code, revision 25ce8063},
  url = {https://github.com/sgl-project/sglang/blob/25ce8063f7dbe444c4e24c562f31f86268e05a23/python/sglang/srt/arg_groups/kv_cache_hook.py},
  note = {Accessed 2026-10-31}
}

@misc{sglang-unified-blog,
  author = {Yichi Zhang and {SGLang Team}},
  title = {Unified Memory in {SGLang}: 1.9$\times$ as many prefixes retained, 37.9\% less time for returning requests},
  year = {2026},
  howpublished = {SGLang Blog},
  url = {https://www.sglang.io/blog/unified-memory},
  note = {Accessed 2026-10-29}
}

@misc{sharegpt,
  author = {{anon8231489123}},
  title = {{ShareGPT\_Vicuna\_unfiltered}},
  year = {2023},
  howpublished = {Hugging Face dataset},
  url = {https://huggingface.co/datasets/anon8231489123/ShareGPT_Vicuna_unfiltered},
  note = {ShareGPT\_V3\_unfiltered\_cleaned\_split.json}
}

@misc{claude-code-traces,
  author = {{SemiAnalysis}},
  title = {{CC Traces---Weka, With Subagents, v7}},
  year = {2026},
  month = jun,
  howpublished = {Hugging Face dataset},
  url = {https://huggingface.co/datasets/semianalysisai/cc-traces-weka-062126},
  note = {June 21, 2026 release}
}

@misc{hicache,
  author = {Zhiqiang Xie},
  title = {{SGLang HiCache}: Fast Hierarchical {KV} Caching with Your Favorite Storage Backends},
  year = {2025},
  month = sep,
  howpublished = {LMSYS Org Blog},
  url = {https://www.lmsys.org/blog/2025-09-10-sglang-hicache/}
}

@misc{cuda-vmm,
  author = {{NVIDIA}},
  title = {{CUDA} Driver {API}: Virtual Memory Management},
  year = {2026},
  url = {https://docs.nvidia.com/cuda/cuda-driver-api/group__CUDA__VA.html},
  note = {Accessed October 30, 2026}
}

@misc{cuda-memcpy-batch,
  author = {{NVIDIA}},
  title = {{CUDA} Runtime {API}: Memory Management},
  year = {2026},
  url = {https://docs.nvidia.com/cuda/cuda-runtime-api/group__CUDART__MEMORY.html},
  note = {cudaMemcpyBatchAsync; accessed October 30, 2026}
}

@misc{qsa,
      title={On the Design of {Qwen3.8-Next} Architecture: Evaluation, Efficiency, and Training Stability},
      author={Zihan Qiu and Zekun Wang and Xiao Li and Yanpeng Li and others},
      year={2026},
      eprint={2608.30320},
      archivePrefix={arXiv},
      url={https://arxiv.org/abs/2608.30320},
}
